\documentclass[11pt,a4paper]{article}

\usepackage[includeheadfoot,
            marginratio={1:1,2:3}, 
          width=412pt, 
            height=688pt,]{geometry} 

\usepackage{amsmath,amsfonts,amssymb,mathrsfs}
\usepackage{systeme}
\usepackage{dsfont}
\usepackage{graphicx}
\usepackage{epsfig}
\usepackage{mathtools}
\usepackage{makeidx}
\usepackage{sectsty}
\usepackage{caption}
\usepackage{etoolbox}
\usepackage{authblk} 

\usepackage{color} 
\usepackage[colorlinks,filecolor=blue,citecolor=blue,unicode]{hyperref}

\definecolor{verde}{cmyk}{.83,.21,1,.08}
\definecolor{darkorchid}{rgb}{0.6, 0.2, 0.8}
\definecolor{darkgreen}{rgb}{0,.5,0}

\def\({\left(}
\def\){\right)}
\def\[{\left[}
\def\]{\right]}

\newcommand{\ii}{\mathrm{i}}

\newcommand{\dd}{\mathrm{d}}

\newcommand{\be}{\begin{equation}}
\newcommand{\ee}{\end{equation}}
\newcommand{\bea}{\begin{eqnarray}}
\newcommand{\eea}{\end{eqnarray}}

\newcommand{\la}{\label}

\newcommand{\eqn}[1]{(\ref{#1})}

\allowdisplaybreaks[2]
\numberwithin{equation}{section}

\title{\textbf{$k$-Minkowski-deformation of $U(1)$ gauge theory}}

\title{\bf $\kappa$-Minkowski-deformation of $U(1)$ gauge theory}
\author[1,2]{V. G. Kupriyanov}
\author[3,4]{M. Kurkov}
\author[3,4]{P. Vitale}
\affil[ ]{}
\affil[1]{\textit{\footnotesize CMCC-Universidade Federal do ABC, 09210-580, Santo Andr\'e, SP, 
Brazil. }}
\affil[2]{\textit{\footnotesize Phisics Department, Tomsk State University, 634050, Tomsk, Russia}}
\affil[3]{\textit{\footnotesize INFN-Sezione di Napoli, Complesso Universitario di Monte S. Angelo Edificio 6, via Cintia, 80126 Napoli, Italy.}}
\affil[4]{\textit{\footnotesize Dipartimento di Fisica ``E. Pancini'', Universit\`a di Napoli Federico II, Complesso Universitario di Monte S. Angelo Edificio 6, via Cintia, 80126 Napoli, Italy.}}
\affil[ ]{}
\affil[ ]{\footnotesize e-mail: \texttt{vladislav.kupriyanov@gmail.com, kurkov@na.infn.it, patrizia.vitale@na.infn.it}}

\begin{document}


\maketitle

\begin{abstract}
\baselineskip=12pt
\noindent 

We construct a noncommutative kappa-Minkowski deformation of  U(1) gauge theory, following a general approach, recently proposed in JHEP 2008 (2020) 041. We obtain an exact (all orders in the non-commutativity parameter) expression for both the deformed gauge transformations and the deformed field strength, which is covariant under these transformations. The corresponding Yang-Mills Lagrangian is gauge covariant and reproduces the Maxwell Lagrangian in the commutative limit.   Gauge invariance of the action functional  requires a non-trivial integration measure which, in the commutative limit, does not reduce to the trivial one.
We discuss the physical meaning of such a nontrivial commutative limit, relating it to a nontrivial space-time curvature of the undeformed theory. 
Moreover, we propose a rescaled kappa-Minkowski noncommutative structure, which exhibits  a standard flat commutative limit.
\end{abstract}



\newpage
\section{Introduction}
Noncommutative Geometry has been the object of a significant research activity in the past years, both from the mathematical and physical point of view, with mainly two objectives from the physics side:  provide  effective models of space-time quantisation and build consistent gauge theories of fundamental interactions.  Both the efforts have been confronted with the issue of a good ``commutative limit'' namely with the idea that  preferred models of space-time and quantum fields should be those which reproduce  the standard  picture of quantum field theory on pseudo-Riemannian manifolds, when the noncommutativity parameter is removed.  

In order to fulfil the goal, we follow here the approach proposed in \cite{BBKL, KV20} where the strategy is to    modify the very definition of  gauge fields and gauge transformations. We choose to work with a specific    space-time noncommutativity,   described by the $k$-Minkowski algebra, 
\begin{equation}\label{kappa}
[\hat x^i,\hat x^j]=\ii\,\Theta^{ij}(\hat x)\,,\qquad\mbox{with} \qquad\Theta^{ij}(x)= 2 \left(a^i x^j-a^j x^i\right)\,,
\end{equation}
 $a^i$  dimensional parameters, 
 and consider specifically pure gauge theories (no matter fields) with gauge group $U(1)$.
 
$\kappa$-Minkowski noncommutativity has attracted much  attention of physicists and mathematicians in various contexts, from tests  of Lorentz symmetry violation and  its relation with  $\kappa$-Poincar\'e quantum groups (see for instance \cite{AmelinoCamelia:1999pm}-\cite{LMM20} and refs. therein) to its applications within gauge and field theory (\cite{Wess}-\cite{Dimitrijevic:2011jg} and refs. therein), not to forget its more formal aspects (\cite{Lukierski}-\cite{Pachol:2015qia} and refs. therein).  Besides its interesting features as a candidate  quantum space-time,    it possesses significant computational  advantages. It is one of those few examples of a non-constant Poisson structure where one may perform an explicit all orders in $\Theta$ calculation of the corresponding star product \cite{Meljanac:2007xb}. 

  In a previous work \cite{KV20} for the case of $\mathfrak{su}(2)$-like non-commutativity we were able to construct an explicit expression for both the non-commutative deformation of the $U(1)$ gauge transformations and the corresponding non-commutative field strength. One of the main goals of the present research is to adopt the same approach for $\kappa$-Minkowski space-time.
 
It is known that, for generic coordinate dependence of $\Theta$, where 
one needs to employ the general Kontsevich star product \cite{Kontsevich}, 
\begin{equation}
 \label{Kon}
f\star g=f\cdot g+\frac{i}{2}\,\Theta^{ij}(x)\,\partial_i f\partial_j g+\dots\,,
\end{equation}
 ordinary derivations violate  Leibniz rule,
\begin{equation*}
\partial_k(f\star g)=(\partial_k f)\star g+f\star(\partial_k g)+\frac{\ii}{2}\,\partial_k\Theta^{ij}(x)\,\partial_i f\,\partial_j g+\dots\,
\end{equation*}
whereas twisted or star derivations, although giving rise to a well defined differential calculus,  might not   reproduce the correct commutative limit.   
Thus,  instead of looking for a deformed differential calculus, we shall deform the very definition of  $U(1)$ gauge transformations, 
\begin{equation}
\label{gautra}
\delta^0_f A\to\delta_f A =\partial f+\dots
\end{equation}
in such a way that the algebra of gauge parameters closes with respect to the star commutator, namely 
\begin{equation}
[\delta_f,\delta_g]A=\delta_{-i[f,g]_\star}A\,.\label{gas}
\end{equation}
We shall work in the  slowly varying, but not necessarily small fields. Therefore we discard  higher derivatives terms in the star commutator and take, 
\begin{equation}
-i[f,g]_\star\approx \{f,g\}=\Theta^{ij}(x) \,\partial_i f\,\partial_j g\,.\label{1}
\end{equation}
In the approximation which we have chosen Eq. (\ref{gas}) becomes
\begin{equation}
[\delta_f,\delta_g]A=\delta_{\{f,g\}}A\,,\label{ga}
\end{equation}
and we look for  gauge transformations in the form  (\ref{gautra}) which be compatible with the latter. 

Before proceeding further, let us  notice  that, in the chosen approximation, the algebra of smooth functions defined on space-time is still a commutative one, the product being  the usual point-wise product, but space-time  geometry is deformed, because it acquires a non-trivial Poisson bracket. Therefore one should rather talk about commutative field theory on Poisson-deformed space-time. This doesn't mean however that the gauge theory is first order in the deformation parameter. We shall see that gauge transformations and fields contain all order deformations in $\Theta$. Once such a distinction made, in the following we shall refer to the latter as noncommutative space-time without any further specification, unless otherwise stated.

The paper is organised as follows. In Sect.  \ref{ncg} we find the deformed gauge transformations adapted to $\kappa$ non-commutativity and derive 
in Sect.  \ref{ncfs} the non-commutative field strength. In Sect. \ref{ncym} we introduce the Yang-Mills Lagrangian and  analyse the problem of defining a gauge invariant action, by deforming the integration measure.  Then, Sect. \ref{imcp} deals with the physical interpretation of the non-trivial integration measure which survives the commutative limit, as  a curvature of space-time. Finally, in Sect.  \ref{comlim}, we propose a different approach to the commutative limit, by modifying the non-commutative parameter, and we conclude in Sect.  \ref{concl} with a short summary of our results. 

\section{Non-commutative gauge transformations}\label{ncg}

Let us consider   noncommutative space-time represented by the algebra $\mathcal{A}_\Theta$ with non-constant non-commutativity tensor $\Theta(x)$ linear in $x$ given in (\ref{kappa}). 
Eventually,  
\begin{equation}
\Theta^{ij}(x)=2\,f^{ij}_k\, x^k=2\left(a^i\delta^j_k-a^j\delta^i_k\right)x^k\,\label{Theta}
\end{equation}
shall describe  the $k$-Minkowski space-time  in $n$-dimensions.
According to \cite{KV20,kup-durham},  we  look for  gauge transformations in the form,
\begin{equation}
\label{h2}
\delta_f A_j= \gamma^k_j(A)\partial_kf+\{A_j, f\}\,.
\end{equation}
The latter close the algebra (\ref{ga}) if the matrix $\gamma(A)_a^k$ satisfies the equation,\footnote{The convention used here is: the partial derivative with the upper index is the derivation with respect to the field, $\partial^j_A=\partial/\partial A_j$, while the partial derivative with the lower index is a derivation with respect to coordinate, $\partial_m=\partial/\partial x^m$.}
\begin{equation}
\gamma^l_m\,\partial^m_A\gamma_j^k-\gamma^k_m\,\partial^m_A\gamma^l_j+\Theta^{lm}\,\partial_m\gamma_j^k-\Theta^{km}\,\partial_m\gamma_j^l-\gamma^m_j\,\partial_m\Theta^{lk}=0\,.\label{eq1}
\end{equation}
Let us look  for a perturbative solution 
\be
 \gamma^{k}_j =  \gamma^{k(0)}_j +  \gamma^{k(1)}_j +  \gamma^{k(2)}_j + ...
\ee
 in   $\Theta\sim|a|$, where 
\be
|a|  :=\max{(|a^i|, i=0,..,d-1)} \la{lea}.
\ee 
At t zeroth order we set $ \gamma^{k(0)}_j=\delta^k_j$, to ensure the correct commutative limit. 
At first order we get, 
\begin{equation}
\label{h6}
\partial^l_A{ \gamma^{k(1)}_j}-\partial^k_A{ \gamma^{l(1)}_j}=\partial_j\Theta^{lk}=2\,f^{lk}_j\,,
\end{equation}
yielding 
\begin{equation*}\label{h6a}
 \gamma^{k(1)}_j=-f^k_a(A):=-f_j^{ kl}A_l=-a^kA_j+\delta^k_j (a\cdot A)\,.
\end{equation*}
At second order, following  \cite{Kup27}, we may pose
\begin{equation*}
 \gamma^{k(2)}_j\sim f^k_l(A)f^l_j(A)\,.
\end{equation*}
However, we should observe that,
\begin{equation*}
f^k_l(A)f^l_j(A)=-(a\cdot A) f^k_j(A)\,,
\end{equation*}
meaning that
\begin{equation*}
\left[f^k_l(A)\right]^n=(-1)^{n-1}(a\cdot A)^{n-1}f^k_l(A)\,.
\end{equation*}
That is, the Anzatz for $\gamma$ becomes
\begin{equation}
\gamma^k_j(A)=\delta^k_j\, \alpha(a\cdot A)-a^kA_j\,\beta(a\cdot A)\,, \qquad \alpha(0)=1, \,\,\,\beta(0)=1\, .\label{aG}
\end{equation}
On substituting  in (\ref{eq1}) we find the equation,
\begin{equation}
\beta\,\alpha-z\,\alpha'\,\beta+\alpha'\,\alpha=2\,\alpha\,,\label{ode1}
\end{equation}
where $z=(a\cdot A)$ is the argument of the unknown functions $\alpha$ and $\beta$. In order to get  the simplest  solution satisfying the initial conditions, $\alpha(0)=1$, and $\beta(0)=1$, we just set $\beta(t)=1$, then the eq.  (\ref{ode1}) becomes,
\begin{equation}
\alpha'(\alpha-z)=\alpha\,,\qquad \alpha(0)=1\,.\label{ode2}
\end{equation}
This  is an homogeneous equation with  solution,
\begin{equation}
\alpha(z)=\sqrt{1+z^2}+z\,.
\end{equation}
Thus, we conclude that,
\begin{equation}\label{Gamma}
\gamma^k_j(A)=\delta^k_j \left(\sqrt{1+(a\cdot A)^2}+(a\cdot A)\right) -a^kA_i\,.
\end{equation}
The gauge transformations (\ref{h2}), with $\gamma^k_j(A)$ given  by Eq. (\ref{Gamma})  
and Poisson tensor of $k$-Minkowski type,
close the algebra (\ref{ga}). In the commutative limit, $\Theta\to0$, ($|a|\to0$) they reproduce the standard $U(1)$ gauge transformations.

We note that the constructed in (\ref{Gamma}) solution of the equation (\ref{eq1}) is not unique. For instance one may easily see that the matrix,
\begin{equation*}
\tilde  \gamma^{k(1)}_j= \gamma^{k(1)}_j+s^{kl}_jA_l\,,\qquad s^{kl}_j(x)=s^{lk}_j(x)\,,
\end{equation*}
also satisfies the equation (\ref{h6}) because of the symmetry of the coefficient $s^{kl}_j(x)$ in the two upper indices. This ambiguity will propagate in the higher orders $\gamma^{k(n)}_j$ each of which also can be modified by the corresponding ``symmetric'' term. This fact is related to the redundancies in the L$_\infty$ bootstrap \cite{BBKL} which were discussed in \cite{Blumenhagen:2018shf}. The modified solution of the equation (\ref{eq1}) defines the modified gauge transformations,
\begin{equation}\label{hatA}
\hat \delta_{\hat f}\hat A_j=\gamma^k_j(\hat A)\partial_k\hat f+\{\hat A_j, \hat f\}+s^{kl}_j\,\partial_k \hat f\,\hat A_l+ {\cal O}(\Theta^2)\,
\end{equation}
which close the same gauge algebra, $[\hat \delta_{\hat f},\hat \delta_{\hat g}]\hat A=\hat \delta_{\{\hat f,\hat g\}}\hat A$. Both gauge transformation (\ref{h2}) with the matrix $\gamma^k_j(A)$ given in (\ref{Gamma}) and the one (\ref{hatA}) are solutions of the L$_\infty$ bootstrap with the same initial data. One may check that these two solutions are related by the field redefinition,
\begin{equation*}
\hat A_j(A)=A_j+\frac12\,s^{kl}_jA_k\,A_l+ {\cal O}(\Theta^2)\,,\qquad \hat f(f,A)=f+ {\cal O}(\Theta^2)\,,
\end{equation*}
satisfying the Seiberg-Witten condition \cite{Seiberg:1999vs},
\begin{equation*}
\hat A\left(A+\delta_fA\right)=\hat A(A)+\hat\delta_{\hat f}\hat A(A)\,.
\end{equation*}
The later means that the gauge orbits of the respective models are mapped onto each other. We conclude that constructed in this section solution of the L$_\infty$ bootstrap is not unique, however mathematically equivalent solutions are related by the Seiberg-Witten map.

\section{Non-commutative field strength} \label{ncfs}

In this section, after shortly reviewing  the derivation of the NC field strength in the general case as in \cite{KV20},  we consider the $\kappa$-Minkowski algebra. We are looking for a deformation of the $U(1)$ field strength,
\begin{equation}
{\cal F}_{ij}=\partial_iA_j-\partial_jA_i+{\cal O}(\Theta)\,,\label{F0}
\end{equation}
which should be covariant under  gauge transformations (\ref{h2}), that is,
\begin{equation}
\delta_f {\cal F}_{ij}=\{{\cal F}_{ij},f\}\,.\label{gcc}
\end{equation}
The solution is given by the expression,
\begin{equation}
{\cal F}_{ij}=P_{ij}{}^{kl}\left(A\right)\,\partial_k A_l+R_{ij}{}^{kl}\left(A\right)\,\left\{A_k,A_l\right\}\,,\label{4}
\end{equation}
where
\begin{equation}
P_{ij}{}^{kl}=2\,\gamma^k_mR_{ij}{}^{ml}\,\label{PR}
\end{equation}
and the coefficient function $R_{ij}{}^{kl}\left(A\right)$ should satisfy the equation,
\begin{equation}
\gamma^k_l\,\partial^l_A R_{ij}{}^{mn}+\Theta^{kl}\,\partial_lR_{ij}{}^{mn}+R_{ij}{}^{ml}\,\partial^n_A\gamma^k_l+R_{ij}{}^{ln}\,\partial^m_A\gamma^k_l=0\,.\label{eqR}
\end{equation}

Following  \cite{KV20}  one may construct a  perturbative  solution in $|a|$. The calculation of the first orders suggests  the Ansatz,
\begin{eqnarray}\label{ansatz}
R_{ij}{}^{kl}&=&\left(\delta_i^k\delta_j^l-\delta_i^l\delta_j^k\right)F(z)+ \\
&&\left[A_i\left(a^k\delta_j^l-a^l\delta_j^k\right)-A_j\left(a^k\delta_i^l-a^l\delta_i^k\right)\right]G(z)\,,\notag
\end{eqnarray}
with the initial condition, $F(0)=1/2$ imposed by (\ref{F0}). By substituting   the Ansatz (\ref{ansatz}) in Eq. (\ref{eqR}) the latter becomes,
\begin{eqnarray}
&&a^k\left(\delta_i^l\delta_j^m-\delta_i^m\delta_j^l\right)\left(F^\prime\,(\alpha-z)-2\,F\right)+\\
&&\left[\delta_i^k\left(a^l\delta_j^m-a^m\delta_j^l\right)-\delta_j^k\left(a^l\delta_i^m-a^m\delta_i^l\right)\right]\left(\alpha\,G+\alpha^\prime\,F\right)+  \notag\\
&&a^k\left[A_i\left(a^l\delta_j^m-a^m\delta_j^l\right)-A_j\left(a^l\delta_i^m-a^m\delta_i^l\right)\right]\left(G^\prime\,(\alpha-z)+\alpha^\prime\,G-3\,G\right)=0\,\notag
\end{eqnarray}
which results in a system of three ODEs for the unknown functions $F(z)$ and $G(z)$:
\begin{eqnarray}\label{ODEs}
&&F^\prime\,(\alpha-z)-2\,F=0\,,\\
&&\alpha\,G+\alpha^\prime\,F=0\,,\notag\\
&&G^\prime\,(\alpha-z)+\alpha^\prime\,G-3\,G=0\,.\notag
\end{eqnarray}
One may see however that these equations are not all independent. Indeed, the third equation is a differential consequence of the first two and Eq.  (\ref{ode2}). The solution of the first equation reads,
\begin{equation}\label{F}
F(z)=\frac12\left(\sqrt{1+z^2}+z\right)^2\,.
\end{equation}
Then from the second equation of (\ref{ODEs}) one finds,
\begin{equation}\label{G}
G(z)=-\frac{F}{\sqrt{1+z^2}}=-\frac12\frac{\left(\sqrt{1+z^2}+z\right)^2}{\sqrt{1+z^2}}\,.
\end{equation}
The final expression becomes,
\begin{eqnarray}\label{answer}
R_{ij}{}^{lm}&=&\frac12\left(\delta_i^l\delta_j^m-\delta_i^m\delta_j^l\right)\left(\sqrt{1+z^2}+z\right)^2 \\
&&-\frac12\left[A_i\left(a^l\delta_j^m-a^m\delta_j^l\right)-A_j\left(a^l\delta_i^m-a^m\delta_i^l\right)\right]\frac{\left(\sqrt{1+z^2}+z\right)^2}{\sqrt{1+z^2}}\,.\notag
\end{eqnarray}
As expected it is antisymmetric in both $i$, $j$ and $l$, $m$. 

In order to compute the field strength let us observe that,
 by Eq. (\ref{PR}), we have $P_{ij}{}^{km}=2\,\gamma^k_lR_{ij}{}^{lm}$. Then, after some simplification we find
\begin{eqnarray}
P_{ij}{}^{lm}=2\left(\sqrt{1+z^2}+z\right)R_{ij}{}^{lm} - \frac{\sqrt{1+z^2}+z}{\sqrt{1+z^2}}\, 
a^l\cdot \left(A_i\delta_j^m-A_j\delta_i^m\right)
 \,. 
\end{eqnarray}
and finally,
\begin{eqnarray}\label{Fkappa}
{\cal F}_{ij}&=&\left(\sqrt{1+z^2}+z\right)^3\left(\partial_i A_j-\partial_jA_i\right)+\\
&&\frac{\left(\sqrt{1+z^2}+z\right)^3}{\sqrt{1+z^2}}\left(A_i\partial_j z-A_j\partial_i z\right)
- { 2\left(\sqrt{1+z^2}+z\right)^2 a^k\left(A_i \partial_k A_j - A_j\partial_k A_i\right) }+\notag\\
&&\left(\sqrt{1+z^2}+z\right)^2\left\{A_i,A_j\right\}+\frac{\left(\sqrt{1+z^2}+z\right)^2}{\sqrt{1+z^2}}\left(A_i\,\{A_j,z\}-A_j\,\{A_i,z\}\right)\,.\notag
\end{eqnarray}
Reminding that $z=a\cdot A$ one may easily see that in the commutative limit, $\Theta\to0$, ($|a|\to0$) the above expression reproduce the commutative $U(1)$ field strength 
\be
F^0_{ij}=\partial_i A_j-\partial_jA_i. \la{Fcommut}
\ee 
Moreover (\ref{Fkappa}) transforms covariantly under the gauge transformation (\ref{h2}), i.e., satisfies the condition (\ref{gcc}).

\section{Non-commutative Yang-Mills action}\label{ncym}
Having in hand the NC field strength (\ref{4}) we define the non-commutative Yang-Mills Lagrangian as,
\begin{equation}
 \la{Lagr}
\mathcal{L} = -\frac{1}{4} \eta^{ik}\eta^{jl}\mathcal{F}_{ij}\mathcal{F}_{kl}
\end{equation}
where  $\eta=  \mathrm{diag}(+1,\underbrace{-1,..,-1}_{\mbox{\footnotesize{$d-1$ times}}})$ is the Minkowski metric. 

Upon gauge transformation  (\ref{gcc}) the Lagrangian density (\ref{Lagr})
and its Euclidean version
\be
\mathcal{L} = \frac{1}{4} \delta^{ik}\delta^{jl}\mathcal{F}_{ij}\mathcal{F}_{kl}
 \la{LagrE}
\ee
transform as follows:
\bea
\delta_f{\mathcal{L}} = \left\{\mathcal{L},f\right\}  
= -\partial_{q}\left(\partial_{r}f\cdot\Theta^{rq}\mathcal{L} \right) + \partial_{r}f\cdot\partial_q\Theta^{rq}\cdot\mathcal{L},
\eea
where the last term  is neither vanishing (as it is the case for $\mathfrak{su}(2)$ noncommutativity \cite{KV20}), nor a total derivative, therefore the `naive'  action,
 \be
 S = \int_{\mathbb{R}^d} \dd x\, \mathcal{L} \la{Snaive}
 \ee
  is  \emph{not}  gauge invariant ($ \delta_{f} S\neq 0$). 

In order to overcome the problem we  modify the classical action by introducing a nontrivial weight $\mu(x)$:
\be
S = \int_{\mathbb{R}^d} \dd x\, \mu\,\mathcal{L}. \la{Smu}
\ee
One can easily check that upon gauge transformation the integrand shifts to a total derivative 
\be
\mu\,\delta_f\mathcal{L} =  -\partial_{q}\left(\mu\partial_{r}f\cdot\Theta^{rq}\mathcal{L} \right)
\ee
iff 
\be
\partial_r \left(\mu\cdot \Theta^{qr} \right) =  0. \la{muEq}
\ee
From now on we will address this relation as a ``compatibility condition". Notice that the latter is precisely the condition to be satisfied  for the cyclicity of the star product \cite{feld,gutt}.

\subsection{General solution of the compatibility condition (\ref{muEq}) in arbitrary dimension $d$.}
Substituting
$
\Theta^{ij} = 2\left( a^ix^j - a^jx^i\right)
$
in Eq.~\eqref{muEq} and dividing by $\mu$ we arrive to $d$ coupled linear PDSs for $\ln{\mu}$:
\be
\Theta^{ji}\,\partial_{i}\ln{\mu} =  2 (1-d)\, a^j, \quad\quad j = 0,..,d-1.
\ee

From now on without loss of generality we assume that $a^0\neq 0$\footnote{The choice $a=(1,0,...,0)$ actually corresponds to the standard form of the $k$-Minkowski tensor.}. 
Just two of these equations,
say the 0th and the 1st, are independent. Indeed,
\begin{eqnarray}
(\mbox{Eq.}\, \# \,k) &=& \left(\frac{x^k a^1 - x^1 a^k}{x^0 a^1 - x^1 a^0}\right) \times(\mbox{Eq.}\, \# \,0) \notag\\
& +& \left(\frac{x^0 a^k - x^k a^1}{x^0 a^1 - x^1 a^0}\right) \times (\mbox{Eq.}\, \# \,1) , \quad\quad k=2,...,d-1.
\end{eqnarray}
After a few simple algebraic passages we can represent the  two independent ones in a more elegant way:
\be
\left\{
\begin{array}{l} 
\mathcal{D}_1 \ln \mu  = 1- d \\
\mathcal{D}_2 \ln \mu = 0
\end{array}
\right. \la{D12sys}
\ee
with
\be
\mathcal{D}_1 := x^i\frac{\partial}{\partial x^i}, \quad \quad \mathcal{D}_2 := a^i\frac{\partial}{\partial x^i}. \la{D12}
\ee
Performing the change of variables
\be
\left\{
\begin{array}{l}
x^0 = a^0\cdot w^0 \\
 x^1 = \exp{(w^1)} + a^1\cdot w^0 \\ 
x^k =  w^k\cdot \exp{(w^1)} + a^k\cdot w^0,\quad k\geq 2
\end{array} \right.
\Longleftrightarrow \quad
\left\{
\begin{array}{l}
w^0 = \frac{x^0}{a^0} \\ 
w^1 = \ln{\left(x^1 - \frac{a^1x^0}{a^0}\right)} \\ 
w^k = \frac{a^0 x^k - a^k x^0}{a^0 x^1 - a^1 x^0},\quad k\geq 2
\end{array} \right. ,
 \la{varch}
\ee
we obtain
\be
\mathcal{D}_1  = w^0 \frac{\partial}{\partial w^0} + \frac{\partial}{\partial w^1}, \quad \quad\mathcal{D}_2  =  \frac{\partial}{\partial w^0},
\ee
therefore the \emph{general} solution of the homogeneous system \eqn{D12sys} is given by an arbitrary
function of the remaining $d-2$ \emph{new} variables $w^2$, ..., $w^{d-1}$. 

Combining this result with a particular solution of Eqs. ~(\ref{D12sys})
\be
\ln{\tilde{\mu}} = (1- d)\cdot w^1, \quad\quad \tilde{\mu}(w) \equiv \mu(x(w)),
\ee
we arrive to the general solution of the compatibility condition~(\ref{muEq}):
\be\label{muhigher}
\mu  = \frac{1}{(a^0 x^1  - x^0 a^1)^{d-1}}\cdot F\left(
\frac{a^0 x^2 - a^2 x^0}{a^0 x^1 - a^1 x^0}, \frac{a^0 x^3 - a^3 x^0}{a^0 x^1 - a^1 x^0},..., \frac{a^0 x^{d-1} - a^{d-1} x^0}{a^{0} x^1 - a^1 x^{0}}
\right),
\ee
where $F(w^2,...,w^{d-1})$ is an arbitrary, sufficiently smooth, function of $d-2$ variables.
{The latter expression is considerably simplified if  we  resort to the canonical form of the $k$-Minkowski tensor, $a_0=1, a_i=0$, $ i=1,...,d-1$. This choice is however not the most general possible, since it is precisely the choice of $a$ as a light-like vector which  will produce,  in $d=2$, the correct flat commutative limit, see Sec.~\ref{FlatLimSec}}. 

\subsection{Significance of the commutative limit}
The presence of a non-trivial weight in  the action~(\ref{Smu}) requires a physical clarification of the commutative limit
$
|a| \rightarrow 0$,  unless
\be
\lim_{|a|\rightarrow 0}\mu=1. \la{implim}
\ee
The issue has been  extensively discussed in the literature. For field theories with no gauge interactions the non-trivial measure in the action functional may be absorbed by fields redefinition (see for example  \cite{Dimitrijevic:2003wv} where 
 $\tilde\psi=\mu^{-1/2}\psi$). The same logic is applicable in non-commutative quantum mechanics \cite{Kupriyanov:2012nb}. However the proposed argument cannot be used in gauge theory on $k$-Minkowski space-time \cite{Dimitrijevic:2003pn}. A possible solution to the problem, with the correct commutative limit in non-commutative gauge theory with non-trivial measure, was proposed in \cite{Meyer:2003wj}. The idea is to multiply the Lagrangian by a gauge invariant expression constructed in terms of   covariant coordinates \cite{Wess}, which would compensate the contribution from the measure $\mu$ in the commutative limit. 

Another possibility is to use the twist approach to the construction of noncommutative gauge theory \cite{Dimitrijevic:2011jg}. The latter however, after the application of the SW-map  results in a gauge theory invariant under standard $U(1)$ gauge transformations, which this is not the goal of the present research. 
Finally we mention an interesting approach related to the twisted trace properties of the integral with standard measure (\ref{Snaive}) analyzed in \cite{Durhuus:2011ci}. In recent works \cite{Mathieu:2020ccc} the authors propose a Lagrangian for $k$-Minkowski gauge theory which is twisted gauge covariant. This yields an action  with  standard integration measure (\ref{Snaive}), which  is however gauge invariant only for space-time dimension $d=5$. 

None of the  proposals discussed above is applicable to the problem we are dealing with. Therefore in what follows we describe two possibilities for the physical interpretation of the commutative limit of the proposed model (\ref{Smu}). The first one relates the  measure $\mu(x)$ to the curvature of the target space in the undeformed theory. However, if we wish to reproduce the standard Maxwell theory  of flat space-time as $\Theta\rightarrow 0$ we can modify the original non-commutativity parameter, $\Theta^{ij}\to \tilde\Theta^{ij}=\mu\cdot\Theta^{ij}$, in such a way that the corresponding measure becomes trivial, i.e., $\partial_i\tilde\Theta^{ij}=0$. The latter will lead to a modification of both the non-commutative gauge transformations $\tilde\delta_f A_a$ and the non-commutative field strength $\tilde{\cal F}$.

\section{The integration measure as a curvature of  space-time}\label{imcp}
For   space-time dimension $d\neq4$, we may set 
\be
g_{ij} = \mu^{\frac{2}{d-4}}\eta_{ij}, \la{metriGen}
\quad\quad \big(\mbox{or}\quad g_{ij} = \mu^{\frac{2}{d-4}}\delta_{ij}\quad\mbox{for the Euclidean case}\big)
\ee
so that   the commutative limit of the expression~\eqref{Smu} may be interpreted as the Maxwell action  in a {curved} space-time
\be
\lim_{|a|\rightarrow 0}S =  \int_{\mathbb{R}^d} \dd x\, \sqrt{|g|}\left(- \frac{1}{4}g^{ij} g^{kl}F^0_{ik}F^0_{jl}\right),   \la{Slim}
\ee
\footnote{One has to cancel the overall minus sign in Eq.~\eqref{Slim}, in the Euclidean case. }where, we remind, $F^0$ stands for the commutative field-strength tensor \eqref{Fcommut}.
As for the case $d=4$, which is not allowed in the rescaling \eqn{metriGen}, we recall that the latter  is the only space-time dimension for which    Maxwell action is conformally invariant. This explains why the weight $\mu$ cannot be reabsorbed in a conformal transformation of the metric, as in Eq. (\ref{metriGen}), when $d=4$.

In the remainder of the section  we discuss the case $d=2$ in detail.


\subsection{The $d=2$ case: general discussion}
At $d=2$ the arbitrary function $F$ in the general solution \eqref{muhigher} reduces to a constant, whenever $\mu$ is nonsingular.  
The singularity located on the line $a^0x^1 = a^1x^0$ cuts the $x^0-x^1$ plane, and the values of the constant in the two half-planes 
are in general
different.  
On multiplying and dividing Eq.~\eqref{muEq} by 
$|a|$, we get
\be
\mu = 
 \left\{
\begin{array}{c}
\mathcal{C}_1\cdot (n^0x^1-n^1x^0)^{-1},\quad n^0x^1-n^1x^0 > 0 \\ 
\mathcal{C}_2\cdot (n^0x^1-n^1x^0)^{-1},\quad n^0x^1-n^1x^0 < 0.
\end{array} \right. \la{2dmuB}
\ee
where $\mathcal{C}_{1,2}$ stand for arbitrary constants\footnote{Strictly speaking the constants $\mathcal{C}_{1,2}$ can depend on $a$, and one has to require that the commutative limits of $\mathcal{C}_{1,2}$ are finite and different from zero. Such a dependence, however, does not alter our conclusions, therefore from now on it will be ignored.}  of the dimension of the length, and 
\be
n^{k} := a^{k}/|a|. \la{ndef}
\ee
It is remarkable that the measure~(\ref{2dmuB}) is not compatible with Eq.~(\ref{implim}). 

In order to have a meaningful interpretation of the commutative limit in the sense described above,  we require that
$\mu$ be positive almost everywhere. Therefore we impose that  $\mathcal{C}_{1} > 0$ and $\mathcal{C}_2 < 0$.
Without substantial loss of generality, we set
\be
\mathcal{C}_1 :=\mathcal{C},\quad\mathcal{C}_2 := -\,\mathcal{C},
\ee
where $\mathcal{C}$ is an arbitrary positive constant, so that the measure turns into
\be
\mu = \mathcal{C}\cdot |n^0x^1-n^1x^0|^{-1}. \la{mufinal}
\ee
Let us address the meaning of the singularity at $n^0x^1-n^1x^0 = 0$. To this,  let us take  
 a closer look at the commutative limit. 
As we noticed  above, for  vanishing $|a|$ our action reduces to the Maxwell action~(\ref{Slim}) on the (pseudo) Riemannian
manifold, equipped with the metric tensor
\be
g_{ij} = \mu^{-1} \eta_{ij}, \quad\quad \big(\mbox{or}\quad g_{ij} = \mu^{-1} \delta_{ij}\quad\mbox{in the Euclidean case}\big)
\ee
with  $\mu$ defined by Eq.~\eqref{mufinal}. In order to understand whether this space-time is curved or not, we analyse the scalar curvature\footnote{We remind that a two-dimensional  manifold is curved iff its scalar curvature is different from zero.}: 
\bea
\mathcal{R} 
                   =\pm  (n_j n^j )    \cdot\mathcal{C}^{-2}\mu^{3} \la{RicciCurvLor}
\eea
where the plus and the minus signs correspond to the Lorentzian and Euclidean situations respectively, and
 the factor $n_j n^j$ equals to
\be
n^i n^j \eta_{ij}  = (n^0)^2 - (n^1)^2,  \quad\big(\mbox{or}\quad  n^i n^j \delta_{ij}  = (n^0)^2 + (n^1)^2\quad\mbox{in the Euclidean case}\big).
\ee
If  $n_in^j \neq 0$ then in both Lorentzian and Euclidean cases we have a curved space-time. The scalar curvature is singular on the line
$n^0x^1-n^1x^0 = 0$, therefore the singularity of the measure~\eqref{mufinal} is physical. We are dealing with a geometric singularity of  space-time.

In the Euclidean scenario all vectors are space-like, therefore the commutative limit always corresponds to a curved space. In the Lorentzian situation a curved space-time appears if $|a^1| > |a^{0}|$, i.e. $n^i$ is a \emph{space-like} vector, or if  $|a^0| > |a^{1}|$, when $n^i$ is a \emph{time-like} vector.

\subsubsection{\label{FlatLimSec} The $d=2$ case:  flat commutative limit.}
Throughout this section we use the notations $x := x^1$, $t:=x^0$.
In contrast with the Euclidean model, the Lorentzian one exhibits one more, probably the most interesting option.
The vector $n^i$ can be \emph{light-like}, i.e. $n_in^j = 0$, what corresponds to $a^0 = \pm a^1$. Indeed, for this kind of non-commutativity the scalar curvature~(\ref{RicciCurvLor}) vanishes identically, what implies that we have a flat space-time!

In particular, the pole of the volume factor is a singularity of our coordinate system (see the remark below), whilst no  physical singularity is there. Let us fix  $n^0 = 1=n^1$, $\mathcal{C} = 1$.
The fact that the metric tensor 
\be
g_{ij} = |x-t|\cdot \eta_{ij}
\ee
is not constant, implies  that  the coordinates we are using are  \emph{curvilinear} coordinates on a flat plane.
\begin{figure}[t]
\epsfxsize=2.8 in
\bigskip
\centerline{\epsffile{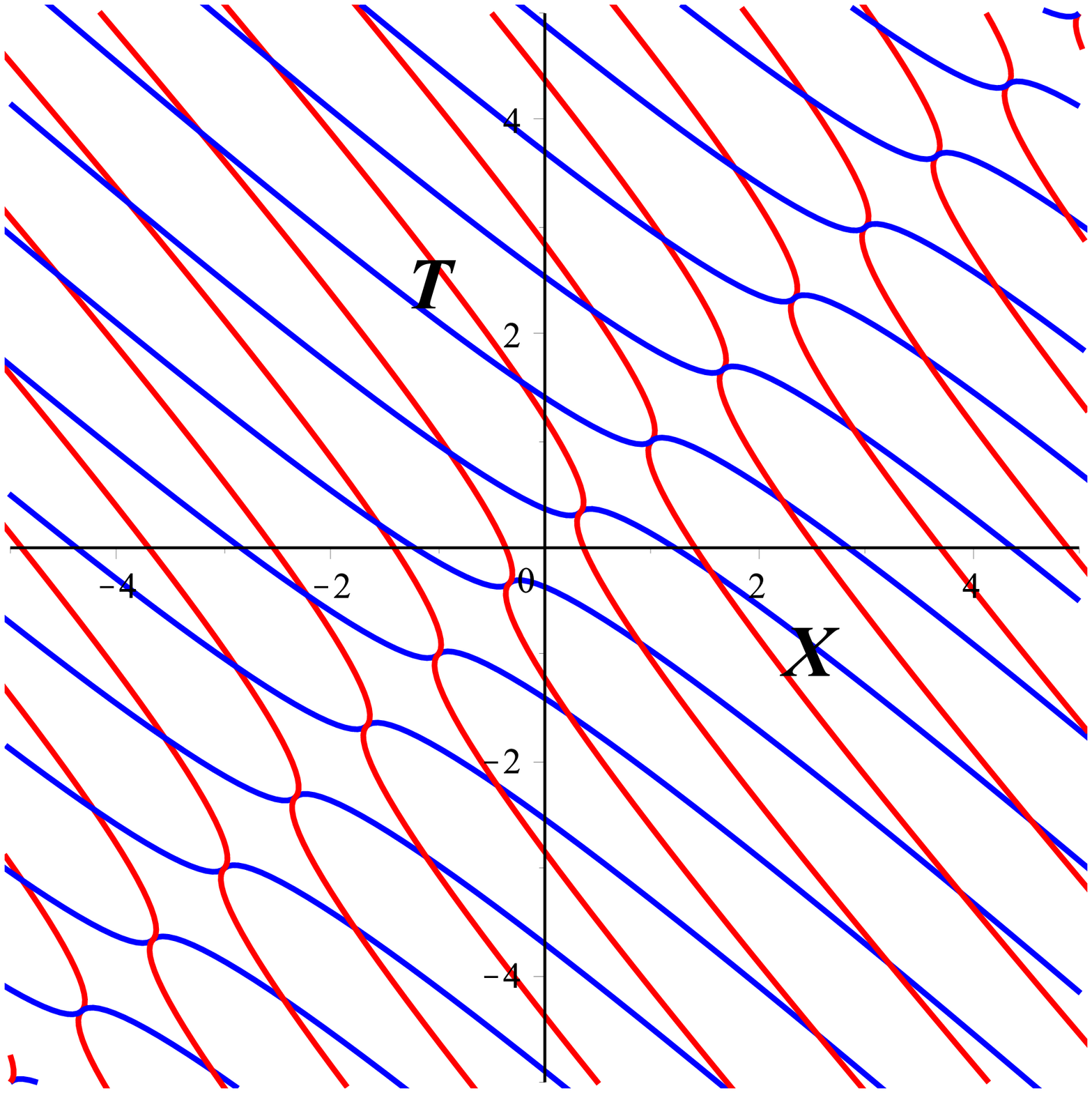}\epsfxsize=2.8 in\epsffile{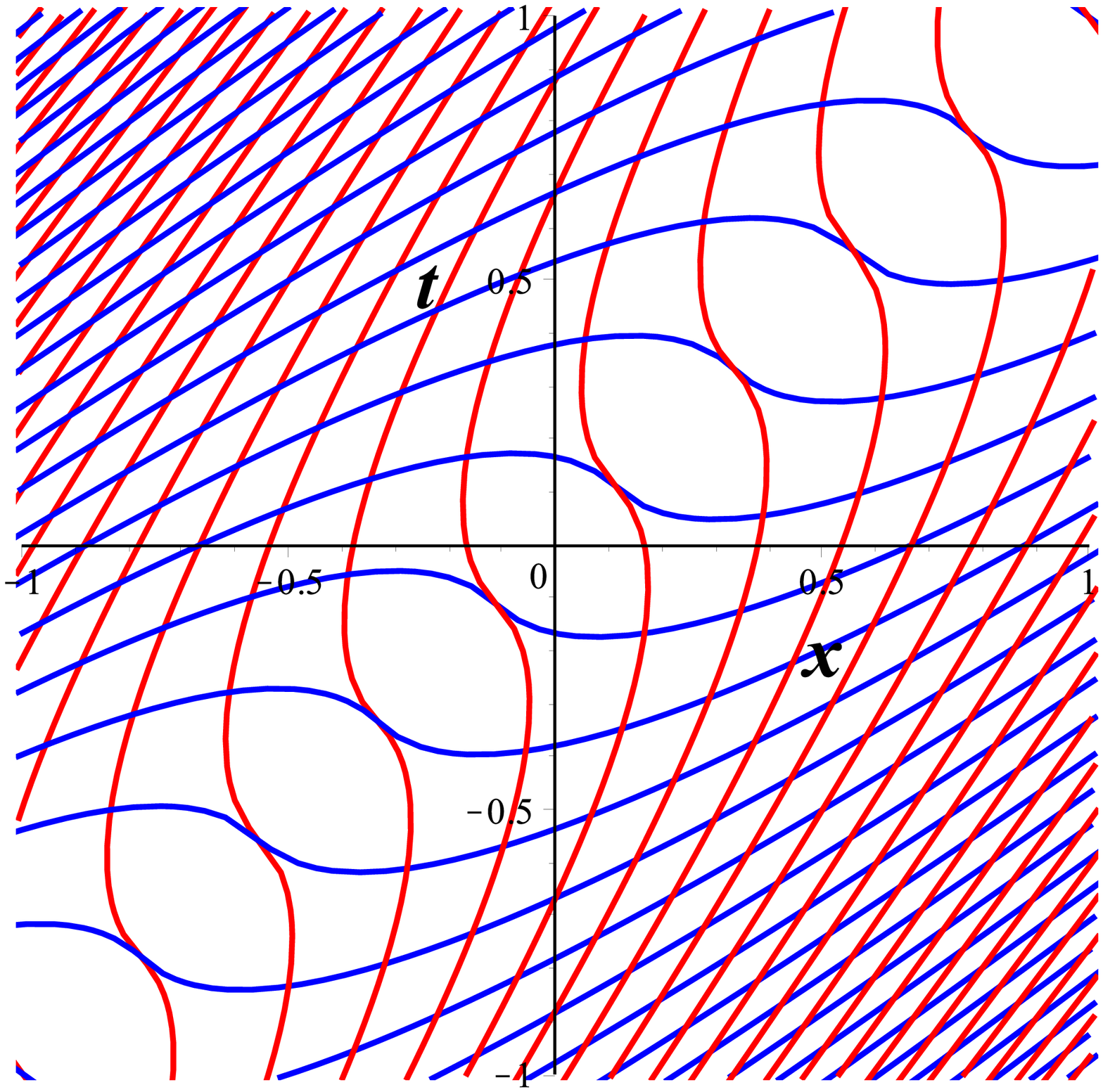}}
\caption{\sl On the left: lines of constant $x$ and $t$ are depicted on the $X-T$-plane in red and blue colours correspondingly.
Fixing $x$ or $t$ in Eq.~(\ref{chvar}) we obtain parametric equations of these lines, where the remaining variable ($t$ or $x$ respectively) plays the role of a parameter.  
On the right: lines of  constant $X$ and $T$ are presented on the $x-t$-plane in red and blue respectively.
Fixing $X$ or $T$ in Eq.~(\ref{chvarinv}) we get parametric equations of these curves, where the remaining variable ($T$ or $X$ correspondingly) stands for a parameter. 
\label{pic}}
\end{figure}
Indeed, one can easily see that by performing the change of variables
{\bea
T &=& {-} \frac{1}{2}\,\mathrm{sgn}(x-t)\cdot(x-t)^2  + \frac{1}{4}\,(x+t) := y^0\nonumber\\
X &=& +\frac{1}{2}\,\mathrm{sgn}(x-t)\cdot(x-t)^2  + \frac{1}{4}\,(x+t) :=y^1, \la{chvar}
\eea}
we convert the metric to the canonical expression
{\be
\bar{g}_{k l}(y) = \frac{\partial x^i}{\partial y^{k}}  \frac{\partial x^j}{\partial y^{l}}\, g_{ij} = \eta_{\alpha\beta},
\ee}
so the limiting action~(\ref{Slim}) takes the standard form
\be
\lim_{|a|\rightarrow 0}S =  \int_{\mathbb{R}^2} \dd y \left(-\frac{1}{4}\eta^{i j} \eta^{k l}\bar{F}^0_{i k}\bar{F}^0_{j l}\right),
\ee 
with
\be
\bar{F}^0_{k l}(y) = \frac{\partial x^i}{\partial y^{k}}  \frac{\partial x^j}{\partial y^{l}}\, F^0_{ij}(x). 
\ee
\\
\noindent{\bf Remark.} The change of variables (\ref{chvar}) is globally invertible,
\bea
t = X  {+}T - \frac{1}{2}\mathrm{sgn}(X {-}T)\cdot \sqrt{|X {-}T|}, \nonumber\\ 
x = X  {+}T + \frac{1}{2}\mathrm{sgn}(X {-}T)\cdot \sqrt{|X {-}T|} \la{chvarinv}, 
\eea
even though the Jacobian of the transformation~(\ref{chvar}) vanishes at $x-t = 0$. In particular the line $x=t$ is
mapped onto the line $X=T$ and the other way around\footnote{Its worth noticing that the Jacobian of the inverse transformation~\eqref{chvarinv} explodes at $X=T$.}. 


The singularity of the volume factor in Eq.~(\ref{Slim}) at $n^0x^1-n^1x^0$ comes out from  Jacobian's singularity, therefore it is related to a parametrisation and has nothing to do with singularities of geometric quantities (in contrast to the situation of $n_i n^j \neq 0$, discussed above). 

For purely illustrative purposes, we plot, using different colours, lines of the coordinate system $(x,t)$ on the plane $X-T$ and vice versa, see Fig.~\ref{pic}. Each point  of each plane is located at the intersection of a single red and a single blue curve. The singularities of the coordinate system, mentioned above, have a transparent origin: at $T=X$ and $t=x$ the red and blue curves intersect at zero angle.

{ Finally, let us notice that the 
\emph{nonlinear} transformation~(\ref{chvar}) leaves the Poisson structure unchanged:
\be
\bar{\Theta}^{ij}(y) := \frac{\partial y^i}{\partial x^k}\frac{\partial y^{j}}{\partial x^l}\,\Theta^{kl}(x) = \Theta^{ij}(y).
\ee
This implies, in particular, that  the nature of space-time noncommutativity is still  $k$-Minkowski, but  with a trivial integration weight, $\mu$. 
}

\section{\label{comlim}Modifyed $\kappa$-Minkowski space-time}
In order to reproduce the standard Maxwell theory in the commutative limit we may proceed in a different way, by  modifying the original non-commutativity to absorb the non-trivial measure,
\begin{equation}\label{Thetam}
\Theta^{ij}(x)\ \ \to\ \ \tilde\Theta^{ij}(x)=\mu(x)\cdot\Theta^{ij}(x)\,.
\end{equation}
First of all let us observe that the modified non-commutativity parameter is still a Poisson structure, i.e., it satisfies  Jacobi identity. Indeed, because of the compatibility condition \eqn{muEq}, $\partial_i\mu\cdot\Theta^{ij}=-\mu\cdot\partial_i\Theta^{ij}$. Therefore 
\begin{eqnarray}\label{jacthetam}
\tilde\Theta^{il}\,\partial_l\tilde\Theta^{jk}+\tilde\Theta^{kl}\,\partial_l\tilde\Theta^{ij}+\tilde\Theta^{jl}\,\partial_l\tilde\Theta^{ki}&=&
\mu^2\left(\Theta^{il}\,\partial_l\Theta^{jk}+\Theta^{kl}\,\partial_l\Theta^{ij}+\Theta^{jl}\,\partial_l\Theta^{ki}\right)\\
&-&\mu^2\left(\partial_l\Theta^{il}\,\Theta^{jk}+\partial_l\Theta^{kl}\,\Theta^{ij}+\partial_l\Theta^{jl}\,\Theta^{ki}\right)\,\notag
\end{eqnarray}
where the first term on the RHS is  zero because it is  the   Jacobi identity for $\Theta^{ij}$, while the second one  is proportional to
\begin{equation}
a^i\left(a^j\,x^k-a^k\,x^j\right)+a^k\left(a^i\,x^j-a^j\,x^i\right)+a^j\left(a^k\,x^i-a^i\,x^k\right)\,
\end{equation}
which is zero as well.
Therefore 
$\tilde\Theta^{ij}$ is a Poisson bi-vector and we  can define a modified Poisson bracket according to
\begin{equation}
\{f,g\}_{\mathrm m}=\tilde\Theta^{ij}(x) \,\partial_i f\,\partial_j g\,.\label{m1}
\end{equation}
Let us notice that, since $\partial_i\tilde\Theta^{ij}(x)=0$, the modified Poisson bracket (\ref{m1}) of two Schwartz functions integrated with the standard measure vanishes,
\begin{equation}\label{im}
 \int_{\mathbb{R}^d} \dd x\, \{f,g\}_{\mathrm m}=0\,.
\end{equation}
The latter  property will be important to show gauge invariance of the action.

Modification of the original non-commutativity will also induce a modification of the expressions for both the non-commutative gauge transformation $\tilde\delta_f A$ and the field strength $\tilde{\cal F}$. Since the modified non-commutativity $\tilde\Theta^{ij}(x)$ is no longer linear in $x$ the functions $\tilde\gamma_j^i(x,A)$, $\tilde P_{ij}{}^{kl}(x,A)$ and $\tilde R_{ij}{}^{kl}(x,A)$ will have an explicit coordinate dependance. For an arbitrary function $\mu(x)$ of the form  (\ref{muhigher}) one can only provide a perturbative expression for these functions. Following \cite{KV20} we have, 
\begin{eqnarray}
\tilde\gamma^k_i(x,A)&=&\delta^k_i-\frac12\, \partial_i \tilde\Theta^{kj} A_j\label{gpsm}\\
&-&\frac{1}{12}\left(2\,\tilde\Theta^{pm}\partial_i\partial_m\tilde\Theta^{jk}+\partial_i\tilde\Theta^{jm}\partial_m\tilde\Theta^{kp}\right)A_jA_p+{\cal O}(\tilde\Theta^3)\,,\notag\\
\label{R2m}
\tilde R_{ij}{}^{pq}(x,A)&=&\frac12\left(\delta_i^{p}\delta_j^{q}-\delta_i^{q}\delta_j^{p}\right)
\\&+&
\frac14\left(\delta^p_i\,\partial_j\tilde\Theta^{kq}-\delta^q_i\,\partial_j\tilde\Theta^{kp}-\delta^p_j\,\partial_i\tilde\Theta^{kq}+\delta^q_j\,\partial_i\tilde\Theta^{kp}\right)A_k
\notag\\
&+&\left(\frac{1}{12}\delta^p_i\,\tilde\Theta^{nm}\,\partial_j\partial_m\tilde\Theta^{kq}-\frac{1}{12}\delta^q_i\,\tilde\Theta^{nm}\,\partial_j\partial_m\tilde\Theta^{kp}\right.\notag\\&-&\frac{1}{12}\delta^p_j\,\tilde\Theta^{nm}\,\partial_i\partial_m\tilde\Theta^{kq}+\frac{1}{12}\delta^q_j\,\tilde\Theta^{nm}\,\partial_i\partial_m\tilde\Theta^{kp}\notag\\
&+&\frac{1}{12}\delta^p_i\,\partial_j\tilde\Theta^{nm}\,\partial_m\tilde\Theta^{kq}-\frac{1}{12}\delta^q_i\,\partial_j\tilde\Theta^{nm}\,\partial_m\tilde\Theta^{kp}\notag\\&-&\frac{1}{12}\delta^p_j\,\partial_i\tilde\Theta^{nm}\,\partial_m\tilde\Theta^{kq}+\frac{1}{12}\delta^q_j\,\partial_i\tilde\Theta^{nm}\,\partial_m\tilde\Theta^{kp}\notag\\
&+&\left.\frac{1}{8}\partial_i\tilde\Theta^{kp}\,\partial_j\tilde\Theta^{nq}-\frac{1}{8}\partial_i\tilde\Theta^{nq}\,\partial_j\tilde\Theta^{kp}\right)\,A_kA_n+{\cal O}(\tilde\Theta^3)\,,\notag\\
\tilde P_{ij}{}^{pq}(x,A)&=&2\,\tilde\gamma^p_l\tilde R_{ij}{}^{lq}\,.\label{PRm}
\end{eqnarray}
The corresponding non-commutative gauge transformations,
\begin{equation}
\label{m2}
\tilde\delta_f A_j=\tilde \gamma^k_j(x,A)\partial_kf+\{A_j, f\}_{\mathrm m}\,,
\end{equation}
close the algebra, $[\delta_f,\delta_g]A=\delta_{\{f,g\}_{\mathrm m}}A$, and reproduce the standard $U(1)$ gauge transformations $\delta^0_f$ in the commutative limit, since $\lim_{|a|\to0}\tilde\Theta^{ij}=0$. The modified field strength,
\begin{equation}
\tilde{\cal F}_{ij}=\tilde P_{ij}{}^{kl}\left(x,A\right)\,\partial_k A_l+\tilde R_{ij}{}^{kl}\left(x,A\right)\,\left\{A_k,A_l\right\}_{\mathrm m}\,,\label{4m}
\end{equation}
is gauge covariant, $\tilde\delta_f\tilde{\cal F}_{ij}=\{f,\tilde{\cal F}_{ij}\}_{\mathrm m}$, and reproduces the Maxwell field strength $F^0_{ij}$ in the commutative limit. Because of (\ref{im}) the corresponding non-commutative action,  with the standard integration measure,
\begin{equation}
 \la{Lagrm}
 S_{\mathrm m} = \int_{\mathbb{R}^{d}} \dd x\, {\cal L}_{\mathrm m} \,,\qquad{ \cal L}_{\mathrm m} = -\frac{1}{4} \eta^{ik}\eta^{jl}\tilde{\cal F}_{ij}\tilde{\cal F}_{kl}\,,
\end{equation}
is gauge invariant and yields the standard Maxwell action in the commutative limit. The obtained result is valid in any dimension. 

For e specific choices of $\mu(x)$ motivated by  considerations of symmetries or some other physical considerations one may ask about the convergence of the perturbative series (\ref{gpsm}) and (\ref{R2m}) and search for explicit all-order expressions. We leave it as an open problem for the future research.

\section{Conclusions} \label{concl}
In this work we construct a gauge theory with  $k$-Minkowski noncommutativity $\Theta^{ij}(x) = 2\left( a^ix^j - a^jx^i\right)$ within the approach proposed in \cite{BBKL, KV20}. We find a non-commutative deformation of the $U(1)$ gauge transformations $\delta_f A$ and of the field strength ${\cal F}$. The latter is  covariant under gauge transformations and reproduces the $U(1)$ field strength in the commutative limit, $|a|\to0$. In order to have a well defined    gauge invariant action, $S=\int {\cal F}^2$, we have introduced a  non-trivial integration measure $\mu(x)$ satisfying $\partial_i(\mu\cdot\Theta^{ij})=0$. In the commutative limit this measure does not tend to a constant, $\lim_{|a|\to0}\mu(x)\neq1$,  which means that the resulting non-commutative gauge theory does not reproduce the standard Maxwell theory. An interpretation of the obtained result is discussed in detail and it relates the non-trivial commutative limit of the measure to the possibility that the original undeformed theory be defined on  curved space.  In  two space-time dimensions with Lorentzian signature we have shown that the limit $|a|\to0$ yields indeed a flat space-time, the original coordinates  just being  \emph{curvilinear} coordinates.

Finally we have proposed a different approach to the commutative limit, which amounts to a redefinition of $\Theta^{\mu\nu}$ in such a way to absorb the non-trivial integration measure. This of course modifies $\kappa$-Minkowski non-commutativity, affecting both gauge transformations and field strength. We have shown that the modification may be carried through consistently, including  the definition of a gauge invariant action with standard commutative limit. The explicit form of gauge fields depends however on a specific choice of the integration measure, which should be made on the basis of physical requests.  We leave it as an open problem which deserves further investigation. 

The present research may be continued in various directions. Based on author's experience, a natural application would be   to extend the analysis performed here  to other types of noncommutativity, e.g., among the linear ones,  angular noncommutativity introduced in ~\cite{DimitrijevicCiric:2018blz}.   Moreover, it would be interesting to study, for the model here considered, the problem of Gribov ambiguity, which affects non-commutative QED on Moyal space, as shown in ~\cite{Canfora:2015nsa}.  We leave these and other open questions for the future research.


\begin{thebibliography}{99}

  \bibitem{BBKL}
  R.~Blumenhagen, I.~Brunner, V.~Kupriyanov and D.~L\"ust,
``Bootstrapping non-commutative gauge theories from L$_\infty$ algebras,''
JHEP \textbf{05} (2018), 097


\bibitem{KV20} 
V.~G.~Kupriyanov and P.~Vitale,
``A novel approach to non-commutative gauge theory,''
JHEP \textbf{08} (2020), 041


\bibitem{AmelinoCamelia:1999pm}
G.~Amelino-Camelia and S.~Majid,
``Waves on noncommutative space-time and gamma-ray bursts,''
Int. J. Mod. Phys. A \textbf{15} (2000), 4301

\bibitem{KowalskiGlikman:2002we}
J.~Kowalski-Glikman and S.~Nowak,
``Doubly special relativity theories as different bases of kappa Poincar\'e algebra,''
Phys. Lett. B \textbf{539} (2002), 126

\bibitem{Lukierski:2002df}
J.~Lukierski and A.~Nowicki,
``Doubly special relativity versus kappa deformation of relativistic kinematics,''
Int. J. Mod. Phys. A \textbf{18} (2003), 7

\bibitem{Borowiec:2009ty}
A.~Borowiec, K.~S.~Gupta, S.~Meljanac and A.~Pachol,
``Constraints on the quantum gravity scale from kappa - Minkowski spacetime,''
EPL \textbf{92} (2010) no.2, 20006


\bibitem{Gubitosi:2013rna}
G.~Gubitosi and F.~Mercati,
``Relative Locality in $\kappa$-Poincar\'e,''
Class. Quant. Grav. \textbf{30} (2013), 145002


\bibitem{aschieri2017} P.~Aschieri, A.~Borowiec and A.~Pacho\l{},
``Observables and dispersion relations in \ensuremath{\kappa}-Minkowski spacetime,''
JHEP \textbf{10} (2017), 152


\bibitem{Meljanac:2016jwk}
S.~Meljanac, D.~Meljanac, F.~Mercati and D.~Pikuti\'c,
``Noncommutative spaces and Poincar\'e symmetry,''
Phys. Lett. B \textbf{766} (2017), 181

\bibitem{LMM20}

F.~Lizzi, M.~Manfredonia, F.~Mercati and T.~Poulain,
``Localization and Reference Frames in $\kappa$-Minkowski Spacetime,''
Phys. Rev. D \textbf{99} (2019) no.8, 085003

F.~Lizzi, M.~Manfredonia and F.~Mercati,
``The momentum spaces of $\kappa$-Minkowski noncommutative spacetime,''
Nucl. Phys. B \textbf{958} (2020), 115117








 \bibitem{Wess}
  J.~Madore, S.~Schraml, P.~Schupp and J.~Wess,
  ``Gauge theory on noncommutative spaces,''
  Eur.\ Phys.\ J.\ C {\bf 16} (2000) 161



\bibitem{Kosinski:1999ix}
P.~Kosinski, J.~Lukierski and P.~Maslanka,
``Local D = 4 field theory on kappa deformed Minkowski space,''
Phys. Rev. D \textbf{62} (2000), 025004



\bibitem{Dimitrijevic:2003wv}
M. Dimitrijevic, L. Jonke, L. Moller, E. Tsouchnika, J. Wess and M. Wohlgenannt, ``Deformed field theory on kappa space-time, Eur. Phys. J. C  \textbf{31}  (2003), 129

\bibitem{Meyer:2003wj}
F.~Meyer and H.~Steinacker,
``Gauge field theory on the E(q)(2) covariant plane,''
Int. J. Mod. Phys. A \textbf{19} (2004), 3349

\bibitem{Dimitrijevic:2003pn}
M.~Dimitrijevic, F.~Meyer, L.~Moller and J.~Wess,
``Gauge theories on the kappa Minkowski space-time,''
Eur. Phys. J. C \textbf{36} (2004), 117


\bibitem{Freidel:2006}
L.~Freidel and E.~R.~Livine,
``3D Quantum Gravity and Effective Noncommutative Quantum Field Theory,''
Bulg. J. Phys. \textbf{33} (2006) no.s1, 111

\bibitem{Arzano:2007ef} 
M.~Arzano and A.~Marcian\`o,
``Fock space, quantum fields and kappa-Poincar\'e symmetries,''
Phys. Rev. D \textbf{76} (2007), 125005


\bibitem{Govindarajan:2008qa}
T.~R.~Govindarajan, K.~S.~Gupta, E.~Harikumar, S.~Meljanac and D.~Meljanac,
``Twisted statistics in kappa-Minkowski spacetime,''
Phys. Rev. D \textbf{77} (2008), 105010



\bibitem{Dimitrijevic:2011jg}
M.~Dimitrijevic and L.~Jonke,
``A Twisted look on kappa-Minkowski: U(1) gauge theory,''
JHEP \textbf{12} (2011), 080.



\bibitem{Lukierski}
J.~Lukierski, H.~Ruegg, A.~Nowicki and V.~N.~Tolstoi,
``Q deformation of Poincar\'e algebra,''
Phys. Lett. B \textbf{264} (1991), 331



\bibitem{Sitarz:1994rh}
A.~Sitarz,
``Noncommutative differential calculus on the kappa Minkowski space,''
Phys. Lett. B \textbf{349} (1995), 42

\bibitem{Agostini:2002de}
A.~Agostini, F.~Lizzi and A.~Zampini,
``Generalized Weyl systems and kappa Minkowski space,''
Mod. Phys. Lett. A \textbf{17} (2002), 2105



 \bibitem{Meljanac:2007xb}
S.~Meljanac, A.~Samsarov, M.~Stojic and K.~S.~Gupta,
``Kappa-Minkowski space-time and the star product realisations,''
Eur. Phys. J. C \textbf{53} (2008), 295-309

\bibitem{Borowiec:2008uj}
A.~Borowiec and A.~Pachol,
``kappa-Minkowski spacetime as the result of Jordanian twist deformation,''
Phys. Rev. D \textbf{79} (2009), 045012

\bibitem{Durhuus:2011ci}
B.~Durhuus and A.~Sitarz,
``Star product realisations of kappa-Minkowski space,''
J. Noncommut. Geom. \textbf{7} (2013), 605

\bibitem{Pachol:2015qia}
A.~Pachol and P.~Vitale,
``$\kappa$-Minkowski star product in any dimension from symplectic realization,''
J. Phys. A \textbf{48} (2015) no.44, 445202




\bibitem{Kontsevich}
  M.~Kontsevich,
  ``Deformation quantization of Poisson manifolds. 1.,''
  Lett.\ Math.\ Phys.\  {\bf 66} (2003) 157
  


  
\bibitem{kup-durham}
  V.~G.~Kupriyanov,
``$L_\infty$-Bootstrap Approach to Non-Commutative Gauge Theories,''
Fortsch. Phys. \textbf{67} (2019) no.8-9, 1910010
  
\bibitem{Kup27}
  V.G.~Kupriyanov,
  ``Non-commutative deformation of Chern-Simons theory,''
  Eur.\ Phys.\ J.\ C {\bf 80} (2020) no.1,  42
  
\bibitem{Blumenhagen:2018shf}
R.~Blumenhagen, M.~Brinkmann, V.~Kupriyanov and M.~Traube,
``On the Uniqueness of L$_\infty$ bootstrap: Quasi-isomorphisms are Seiberg-Witten Maps,''
J. Math. Phys. \textbf{59} (2018) no.12, 123505

 \bibitem{Seiberg:1999vs}
N.~Seiberg and E.~Witten,
``String theory and noncommutative geometry,''
JHEP \textbf{09} (1999), 032
  
 \bibitem{feld} G. Felder and B. Shoikhet, ``Deformation quantisation with traces'', Lett. Math. Phys. {\bf 53}  (2000) 75, [math.QA/0002057].


\bibitem{gutt} S. Gutt and J. Rawnsley, J. Geom. and Phys. {\bf 42} (2002) 12.


\bibitem{Kupriyanov:2012nb}
V.~G.~Kupriyanov,
``A hydrogen atom on curved noncommutative space,''
J. Phys. A \textbf{46} (2013), 245303;\\
V.~G.~Kupriyanov,
``Quantum mechanics with coordinate dependent noncommutativity,''
J. Math. Phys. \textbf{54} (2013), 112105


\bibitem{Mathieu:2020ccc}
P.~Mathieu and J.~C.~Wallet,
``Gauge theories on $\kappa$-Minkowski spaces: twist and modular operators,''
JHEP \textbf{05} (2020), 112;


T.~Poulain and J.~C.~Wallet,
``$\kappa$-Poincar\'e invariant orientable field theories at one-loop,''
JHEP \textbf{01} (2019), 064


T.~Poulain and J.~C.~Wallet,
``$\kappa$-Poincar\'e invariant quantum field theories with KMS weight,''
Phys. Rev. D \textbf{98} (2018) no.2, 025002





  
\bibitem{DimitrijevicCiric:2018blz}
  M.~Dimitrijevic Ciric, N.~Konjik, M.~A.~Kurkov, F.~Lizzi and P.~Vitale,
  ``Noncommutative field theory from angular twist,''
  Phys.\ Rev.\ D {\bf 98} (2018) no.8,  085011

\bibitem{Canfora:2015nsa}
  F.~Canfora, M.~Kurkov, L.~Rosa and P.~Vitale,
  ``The Gribov problem in Noncommutative QED,''
  JHEP {\bf 1601} (2016) 014



\end{thebibliography}
\end{document}